# Generation of Au nanorods by laser ablation in liquid and their further elongation in external magnetic field

G. A. Shafeev*[†], I. I. Rakov, K. O. Ayyyzhy, G. N. Mikhailova, A. V. Troitskii, O.V. Uvarov

A.M. Prokhorov General Physics Institute of the Russian Academy of Sciences, 38, Vavilov street, 119991, Moscow, Russian Federation

[†]Also at: National Research Nuclear University MEPhI (Moscow Engineering Physics Institute), 31, Kashirskoye highway, 115409, Moscow, Russian Federation

*shafeev@kapella.gpi.ru

**Abstract**

Laser-assisted single-step generation of elongated Au NPs is reported. Laser-generated Au NPs have some fraction of spherical NPs and elongated NPs with aspect ratio of 6 - 8. The behavior of these NPs in a permanent magnetic field up to 7 Tesla is experimentally studied using in-situ optical absorption spectrometry. It is found that magnetic field causes irreversible changes in the aspect ratio of elongated Au NPs. Residence in magnetic field for time of order of tens minutes is accompanied by further elongation of Au NPs and formation of Au nanowires with aspect ratios up to 17 – 18. This is corroborated with TEM images of Au NPs before and after the action of magnetic field. The results are interpreted on the basis of interaction of external magnetic field with that of electrons that take part in longitudinal plasmon oscillations in elongated Au NPs.

**Introduction**

Laser ablation of solids in liquids is well-established technique. It enables generating large variety of nanoparticles (NPs). In principle, the so generated NPs are free of any counter-ions and surface active substances that are typical of NPs obtained by chemical methods. Au is one of the most popular target materials owing to its chemical neutrality. Au targets ablated in water provide only $H_2O$ and Au NPs. Others less noble metals may interact with the liquid due to high temperature in the laser-exposed area of the target. Usually, the size of NPs generated by laser ablation in liquids depends on a number of experimental parameters, such as laser fluence on the target, laser wavelength, laser pulse duration, etc. The generated NPs remain in the same liquid and may undergo so called laser fragmentation owing to interaction of individual NPs with laser beam inside the liquid. Fragmentation proceeds via melting of NPs, and typical shape of NPs generated in this way is a sphere. This is stipulated by the surface tension of the melt that governs the shape of NPs to spherical one in order to minimize the surface energy. There are at least two points of view on the process of NPs formation in the laser ablation in liquids. First hypothesis describes the formation of NPs by their condensation from single atoms presented in the plasma plume [1]. In fact, this point of view is not specific to the ablation process in liquid environment and coincides with the same interpretation valid for ablation in vacuum or diluted gases. The difference however is relatively dense medium that is formed around the exposed area of the target in liquid that consists in liquid vapors. The presence of this medium inevitably alters the condensation rate of NPs from atoms that may be presented in this dense environment. Alternative point of view is presented by hydrodynamic interaction of vapors of the liquid with the melt on the target surface during the laser pulse [2]. Indeed, the pressure of vapor of liquid that surrounds the target is of order of several GPa. This pressure splashes the melt from the target surface, so that primary species that appear in the gas-vapor bubble above the target are already NPs. Most probably, both approaches are relevant, and their relative contribution to the NPs formation depends on the experimental conditions of the ablation experiments, such as laser fluence, pulse duration, the height of the liquid layer above the target surface, etc.

Recently we have reported on the influence of external magnetic field onto the process of laser-induced fragmentation of Au NPs [3]. The laser-induced breakdown of liquid on Au NPs is accompanied by laser-produced plasma, which accelerates the process of laser-induced fragmentation. This process consists in the decrease of average size of NPs with exposure time due to their melting and subsequent splitting into smaller fragments. As the result, the size distribution function of NPs shifts to lower sizes. The external magnetic field enhances the fragmentation process due to confinement of plasma in magnetic field. The average temperature of plasma is increased due to higher rate of collisions of electrons. Moreover, the Larmor radius of electrons in the field of order of 10 T is comparable with the diameter of the laser beam waist in the colloidal solution (about 100 μm). Au is a diamagnetic material, so the external magnetic field can hardly influence the morphology of Au NPs.

In our work we used the technique of laser ablation in liquid for the preparation of Au NPs. Typically the shape of Au NPs generated by this way is close to spherical since their formation proceeds through liquid state of the metal. Surface tension of the melt tends minimizing the surface energy that results in spherical shape. Under repeated exposure of NPs to laser radiation their size distribution function shifts to lower sizes, the process which is known as laser fragmentation. Generation of nanowires is not typical for laser ablation in liquids. Usually NPs generated by laser ablation of solids in liquids have spherical shape []. This shape can be expected both from view point of hydrodynamics and from point of view of condensation of atoms form laser plume. From view point of hydrodynamics spherical shape of NPs is due to the surface tension of the melt, and a sphere provides minimal surface energy. From the view point of atoms condensation in laser plume above the target spherical shape can be explained via homogeneous nucleation of atoms into drops. Laser generation of elongated NPs is observed in some specific cases. For example elongated Au NPs are formed under laser ablation in liquid that contains beta-active nuclides, such as Tritium [4]. Elongated shape in this case is attributed to the excessive charge of NPs due to free electrons presented in the liquid owing to beta-decay. Another example in which elongated shape of laser-generated NPs is observed under electrical bias of the target with respect to the liquid volume [5]. Again, formation of elongated NPs is due in this case to the electric charge of the particles. However recently it has been demonstrated that the presence of $Ca^{2+}$ ions in water (at quantity equivalent to drinking water) leads to single-step generation of elongated Au NPs [6].

In this work we show that the permanent magnetic field of several Tesla may affect diamagnetic elongated Au NPs in a liquid without any laser radiation causing their further transformation into nanowires (NWs). This is corroborated with the changes of the extinction spectra of the colloidal solution and Transmission Electron Microscope images of NPs before and after their residence in the magnetic field.

**Experimental**

Laser ablation of a bulk Au target in water was the technique of preparation of initial NPs. The experimental details are given elsewhere [2]. Briefly, Au target of 99.9 % purity was placed on the bottom of a glass cell filled with water purified with reversed osmosis. The water contained small amount of Na and K salts (as chlorides) [6]. For generation of NPs an Ytterbium fiber laser with pulse duration of 80 ns, repetition rate of 20 kHz and pulse energy of 1 mJ at 1060-1070 nm was used. Laser radiation was focused on an Au bulk target by an F-Theta objective. The laser fluence on the target surface amounted to nearly 13 $Jcm^{-2}$ (estimated by the size of heat-affected zone on the target). Laser beam was scanned across the sample surface at the speed of 500 mm/s by means of a galvo-optic mirror system and F-theta objective. The cryogen-free magnetic system with a refrigerator was used as a source of permanent magnetic field. The superconducting solenoid from Nb-Ti alloy wire provided the magnetic field up to 7.5 T. The cell was fixed in the middle part of the magnet where the field is homogeneous. As a light source for spectra acquisition f halogen lamp was used placed at the other end of the magnet. The morphology of Au NPs was analyzed by means of Transmission Electron Microscopy (TEM). Extinction spectra of colloidal solutions were acquired using an Ocean Optics UV-Vis fiber spectrometer in the range of 400 - 900 nm (determined by the spectrum of the bulb used). The setup with magnetic field is shown in Fig. 1.

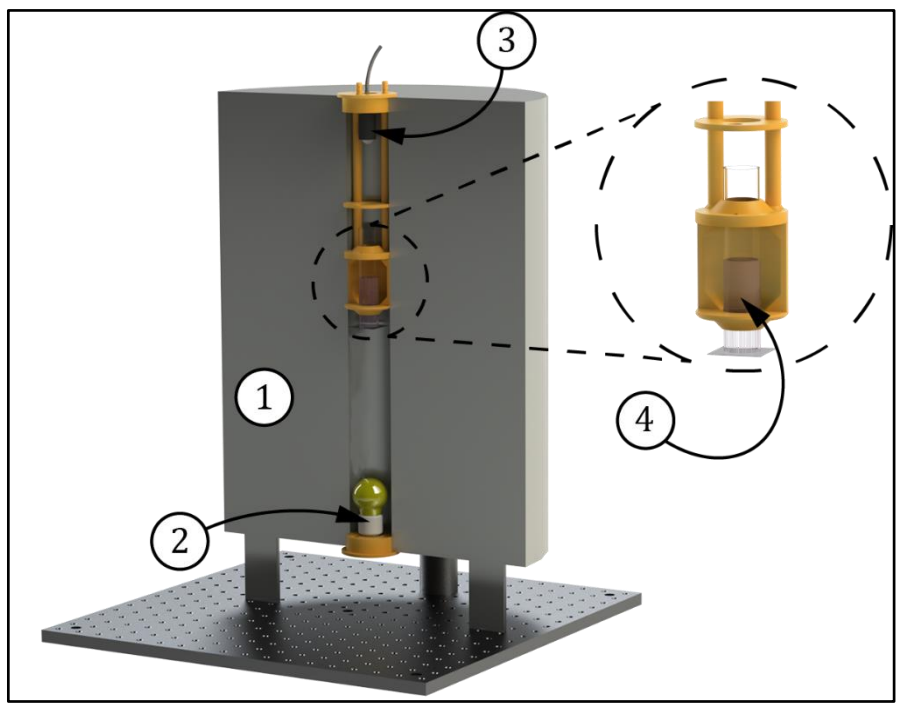

Fig. 1. Experimental setup for measurements of extinction spectra of Au NPs colloidal solution in permanent magnetic field. 1 – superconducting magnet, 2 – bulb, 3 – collimator, 4 – colloidal solution.

The extinction spectra of colloids were taken with respect to pure water in the same cell as a reference.

**Results**

The extinction spectrum of colloidal solutions of Au NPs generated by laser ablation of a bulk Au target is presented in Fig. 1, a. It was found that in the conditions of the present work the extinction spectrum depends on the thickness of the liquid layer above the target. One can see that all spectra are characterized by a wing of OD in red region of spectrum. This wing is due to the presence of elongated Au NPs [7]. The colloidal solution of these NPs looks blue instead of red typical of spherical NPs. Extinction spectrum of elongated Au NPs is characterized by two absorption peaks. One of them is located near 520 nm and corresponds to transverse plasmon resonance. This peak is common for both spherical and elongated Au NPs. The other peak that corresponds to longitudinal plasmon resonance is due to the oscillations of free electrons in the NPs along their longer dimension. The position of this peak depends on the aspect ratio q of elongated Au NPs that is the ratio of their length to their diameter. The higher is q, the more the second peak is shifted to the red region of spectrum. The relative fraction of elongated NPs can be deduced by decomposition of the extinction spectrum of the obtained colloidal solution into peaks that correspond to NPs with aspect ratio q from 1:1 to 1:2 and 1:3. Of course, elongated Au NPs also have the peak at 520 nm.

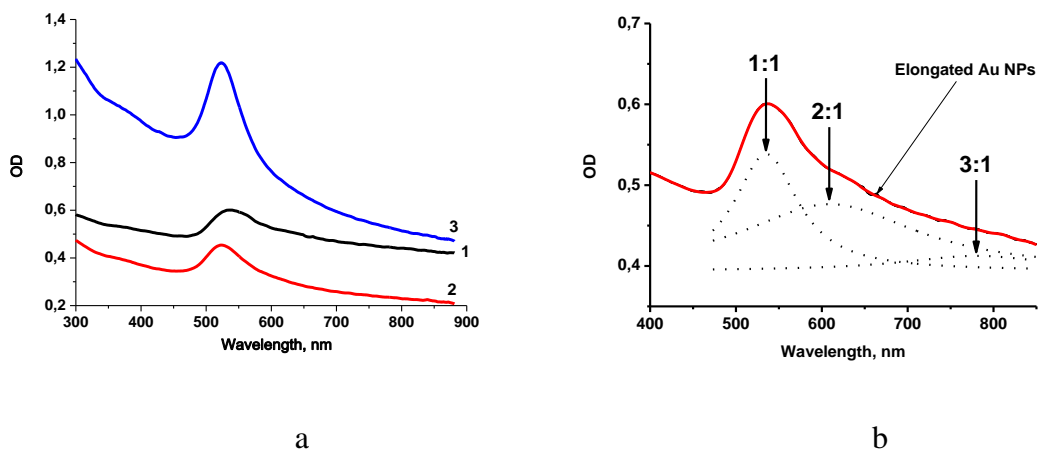

a                                       b

Fig. 2. Extinction spectra of colloidal solutions of Au NPs in water generated in various experimental conditions. Common parameters are: pulse repetition rate of 20 kHz, pulse duration of 80 ns, average power of the laser beam of 20 kHz. 1 – liquid layer thickness of 1.28 mm, exposure time of 1 min, 2 - liquid layer thickness of 2.25 mm, exposure time of 1 min, 3 – liquid layer thickness 2.25 mm, exposure time of 3 min (a). Decomposition of the spectrum 1 in Fig. 2, a into 3 peaks that correspond to spherical NPs (aspect ratio of 1), elongated NPs with aspect ratio of 1:2 and 1:3 (b).

The solution 3 looks red in appearance, while two other solutions look rather blue. Red coloration is typical of spherical Au NPs. As soon as Au NPs have elongated shape they demonstrate so called longitudinal plasmon resonance. The higher is the aspect ratio of Au NPs, the more the peak of this resonance is shifted to the red region of spectrum. The position of the second maximum that corresponds to longitudinal plasmon resonance in Au NPs scales linearly with the aspect ratio q [7]. One may conclude that the increase of the exposure time leads to preferential formation of spherical nanoparticles. This can be explained by the increased probability of secondary exposure of already generated au NPs by laser beam and their eventual fragmentation to spherical NPs. The lower is the thickness of the liquid layer over the target surface the lower is the probability of the second exposure of NPs by the laser beam. Therefore they may preserve their initial shape, which is apparently elongated. There is also a certain influence of the relative size of the exposed surface of the target to the surface of the cell. Wider cells decrease the probability of the second exposure of already generated NPs, since most of the exposure time they remain on the periphery of the cell.

TEM view of Au NPs generated by laser ablation of a bulk Au target is presented in Fig. 3. One can see that the elongated NPs represent a significant part of NPs. In this sense one may declare the generation of Au nanowires.

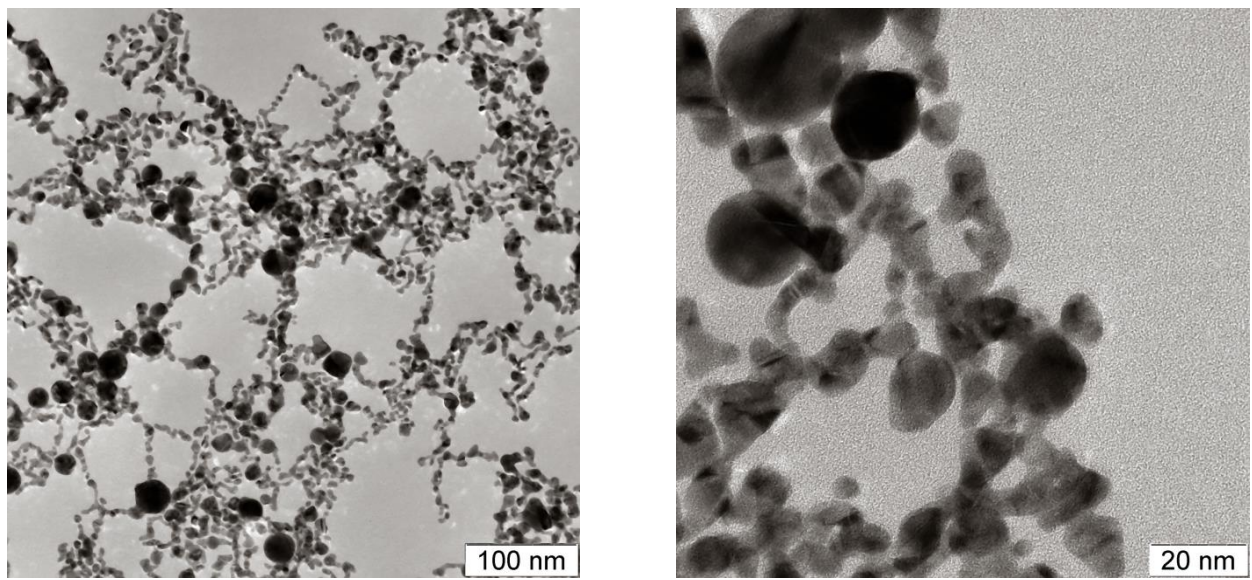

Fig. 3. TEM view of elongated Au NPs generated by laser ablation in water in presence of $Ca^{2+}$ cations. The image corresponds to the spectrum 1 in Fig. 2. Scale bar denotes 100 nm (left) and 20 nm (right).

Size distribution of spherical NPs (aspect ratio of 1) and distribution of aspect ratios for elongated Au NPs are shown in Fig. 4, a and b, as deduced from TEM images. Spherical NPs have average diameter of 20 nm. Aspect ratio of elongated NPs is rather broad with maxima at 4 and 9, though a significant part of elongated NPs is characterized by $q = 2$. The fraction of spherical particles is relatively small.

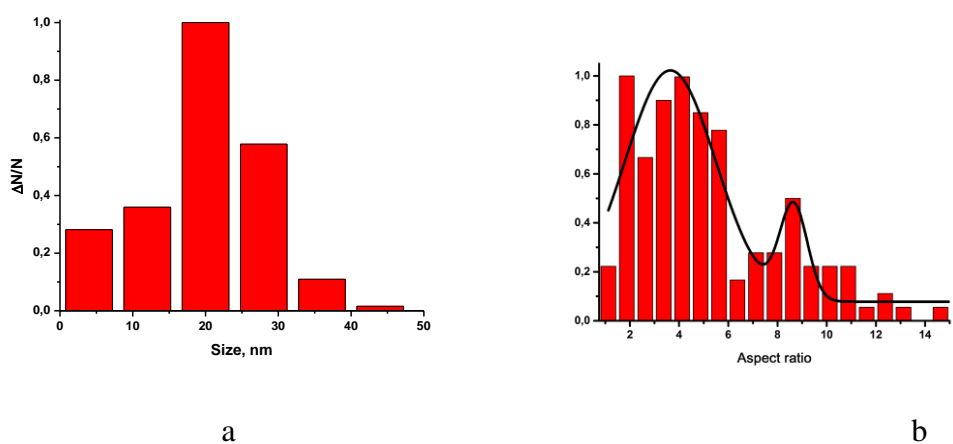

a            b

Fig. 4. Size distribution of spherical NPs (a) in Fig. 3 and several other images, and aspect ratio distribution of elongated Au NPs (b).

One can see that NPs with q = 2 are the most numerous. However, there are a lot of NPs with higher q with maximum at q = 8 - 9.

Laser-generated elongated Au NPs were then placed into magnet in the place where the field is homogeneous. In the first set of experiments, the magnetic field was varied from 0 to 7 T by steps of 0.5 T and back to 0. It is clear that there is certain hysteresis in the variations of extinction spectra in the red region. These variations should be attributed to the elongation of Au NPs into nanowires. However, the time of switching between various values of magnetic field induction is determined by the software of controller of the magnet and is out of control. Irreversible changes in extinction spectrum of the colloidal solution subjected to magnetic field are demonstrated in Fig. 5.

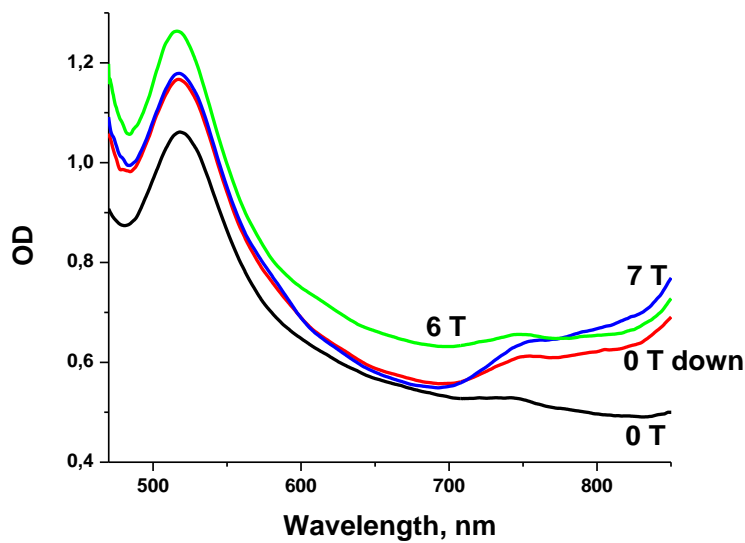

Fig. 5. Optical density of the colloidal solution before magnetic field (0 T), in the field of 6 T, 7T, and 0 T after 7 T. "0 T down" means the decrease of the magnetic field from 7 to 0 Tesla. The cyclic variation of the field strength from 0 to 7 T and back takes about 2 hours.

One can see that under cyclic variation of the magnetic field from 0 to 7 T and back the wing in the red region of spectrum is formed that corresponds to the elongation of the initial elongated Au NPs and remains after decreasing the field to 0 Tesla. However, the residence time of the colloid in the field of various strengths is out of control since it is determined by the software of the field (current) controller. For this reason another set of experiments was carried out with taking extinction spectra at fixed exposure time inside the magnet. If the elongation of NWs proceeds via attachment of smaller NPs, then this process is limited by diffusion of NPs. With field decrease some chains are broken to smaller fragments but not all of them. This can be deduced from the remaining wing in the red region of spectrum. This process requires significant time therefore, the time interval for acquiring spectra should be long enough for diffusion to occur. The first spectrum was taken immediately after placement of the cell inside the magnet (t = 0). The sequence of spectra taken against time of the action of magnetic field on the colloidal solution of Au NPs is shown in Fig. 6.

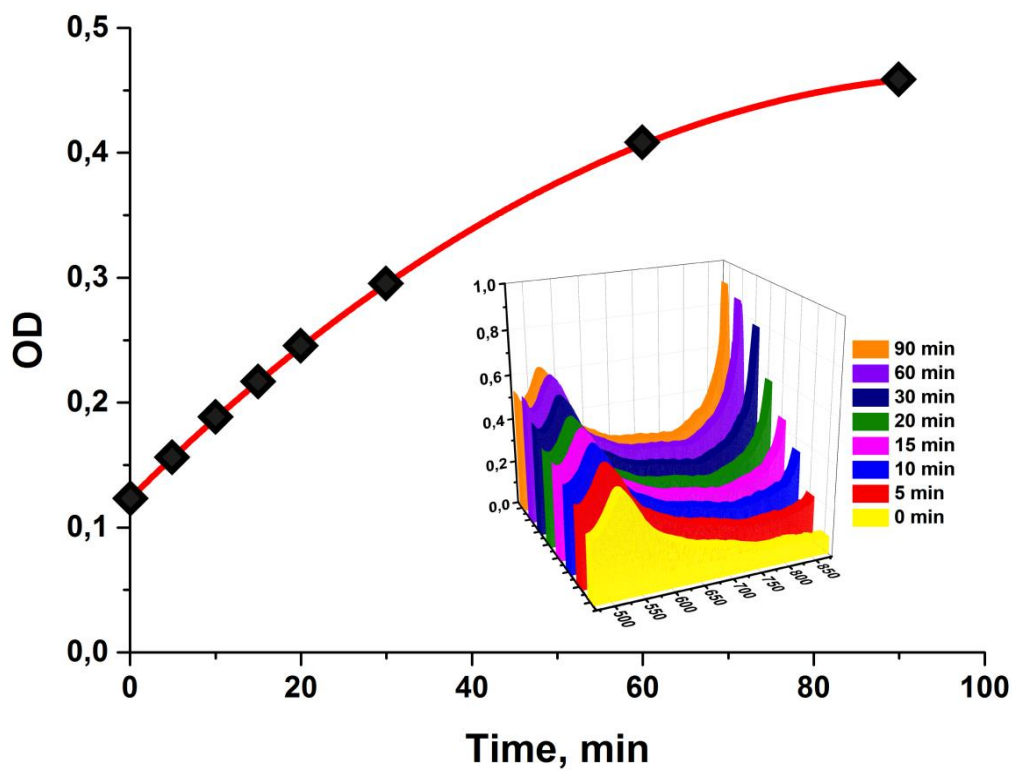

Fig. 6. Optical density of the colloidal solution of elongated Au NPs at 900 nm as the function of time of residence in a permanent magnetic field of 7 Tesla. The inset shows the variation of its extinction spectra in the range 450 – 900 nm for different residence time at 7 T.

One can see that the absorption in the red region of spectrum gradually increases with residual time inside the magnetic field and then starts to saturate in 90 minutes.

TEM images of NPs after magnetic field support this conclusion (Fig. 7). One can see increased fraction of Au NWs after exposure to magnetic field.

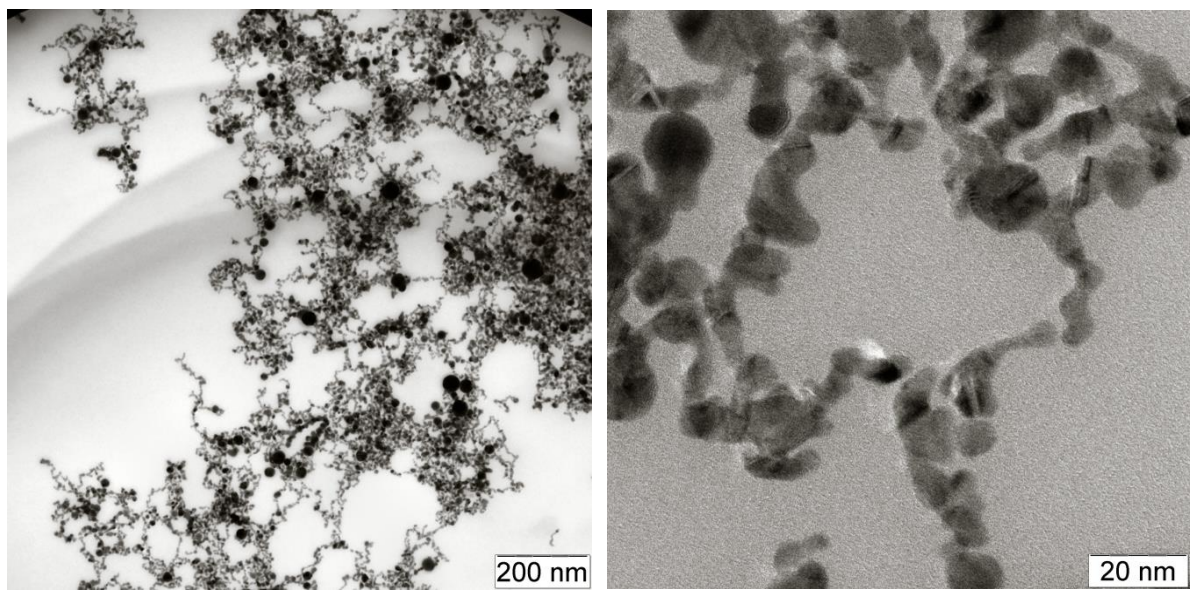

Fig. 7. TEM images of Au NPs after magnetic field of 7 T. Scale bar denotes 200 nm (left) and 20 nm (right).

The magnetic field alters the aspect ratio of NWs in the colloid. Longer wires appear after the action of magnetic field, as shown in Fig. 8.

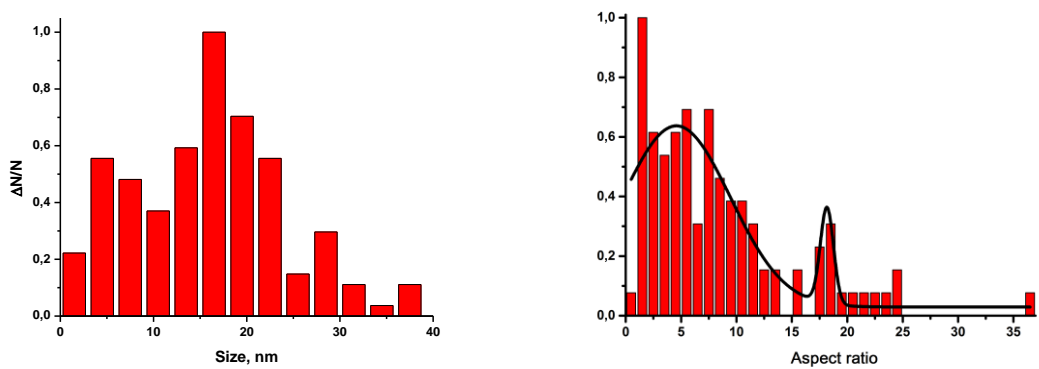

Fig. 8. Size distribution of spherical NPs (left) and aspect ratio of Au NPs subjected to magnetic field of 7 T during 90 minutes (right).

The size distribution of spherical NPs did not change much after exposure to magnetic field. Aspect ratio of elongated Au NPs subjected to exposure in magnetic field is shifted to higher values compared to aspect ratio of initial nanowires (see Fig. 8). New maximum appears at q = 17 – 18. This means that elongated Au NPs are oriented by magnetic field in such a way that they may be connected to each other by their ends. Still, a number of NPs are NPs with q = 2 and spherical particles. It is interesting to note that the size distribution of spherical NPs is not affected by treatment with magnetic field.

**Discussion**

In the conditions of the present work there is no plasma of optical breakdown of the liquid that may interact with magnetic field. The experiments are carried out at room temperature. The heating of the colloidal solution by radiation of the bulb is negligible, since the distance between it and the cell was more than 60 cm. However, the ensemble of elongated Au NPs shows ferromagnetic-like behavior in the magnetic field – the elongated Au NPs are oriented in the field and undergo irreversible changes in their morphology.

Bulk gold is a diamagnetic material as well as Ca-containing compounds are. However, there are a number of communications in which the ferromagnetic behavior of Au nanoparticles (NPs) is reported. Typically, ferromagnetic properties of Au NPs are observed if Au NPs are surrounded by some polar ligands [9]. The effect is specific only for NPs of Au. Ligands, such as thyol group -$SO_2$ are easily adsorbed on the surface of Au causing the shift in electronic density in NPs, especially if they have small size of few nanometers. All the cases of ferromagnetic behavior of Au NPs are related to some environment of NPs that alters the electronic density in NPs. This is concern of Au NPs embedded into Si or Bi matrix [10]. Similar behavior is observed in case of Au clusters deposited onto Si substrate.

In different realizations, Au NPs may exhibit dia-, para-, and ferromagnetic behavior. There are numerous theoretical models that account for these types of magnetism. The easiest case is Au NPs with adsorbed ligands, say, thiol-containing ligands (with $SO_2$ group). The strong covalent bond Au-S depletes the Fermi level of Au. This is especially pronounced in case of small Au NPs (about 2 nm in diameter). This is equivalent to the orbital magnetism and explains remnant magnetization of samples. It should be stressed that Au NPs studied in this work are the purest NPs of Au compared to previously investigated ones [11]. Indeed, the magnetism of Au NPs has been reported for Au NPs embedded into metallic Bi [10]. Formally, these NPs are free though the presence of Bi layer may alter the electronic structure of Au. Typically, ferromagnetic behavior of Au NPs is observed at low temperatures 100 – 200 K). There is remnant magnetization of Au NPs at 0 external magnetic field.

However, the apparent ferromagnetic behavior of elongated Au NPs in this work has nothing to do with their intrinsic magnetic properties. First of all, the samples of elongated Au NPs show no remnant magnetization after the action of magnetic field, unlike in the works mentioned above.

The observed evolution of extinction spectrum can be explained on the basis of interaction of plasmon oscillations of electrons in the magnetic field. This interaction is different with electrons in spherical NP and elongated ones. Plasmon oscillations of electrons are equivalent to an alternative current of very high (optical) frequency. Therefore, this current creates its own magnetic field that interacts with the external magnetic field. Plasmonic oscillations of electrons in spherical NPs are degenerated – all 3 plasmon resonances have the same frequency. The electrons that participate in plasmon oscillations in spherical NPs have no component of mechanical momentum of oscillating electrons on the axis of the NP.

In the elongated NPs longitudinal plasmons may interact more efficiently with external magnetic field since one dimension of the elongated nanoparticle is longer than two others. Electrons that oscillate along long axis may have non-zero momentum with respect to the long axis of the NP. Their own magnetic field oscillates with the frequency of longitudinal plasmon resonance $\omega_{long} \sim 10^{13}$ s$^{-1}$. Magnetic field may interact only with another magnetic field. In our case this second field is due to longitudinal oscillations of electrons in elongated NPs. Fig. 9 shows the relative position of both external magnetic field and the magnetic field induced by longitudinal oscillations of electrons.

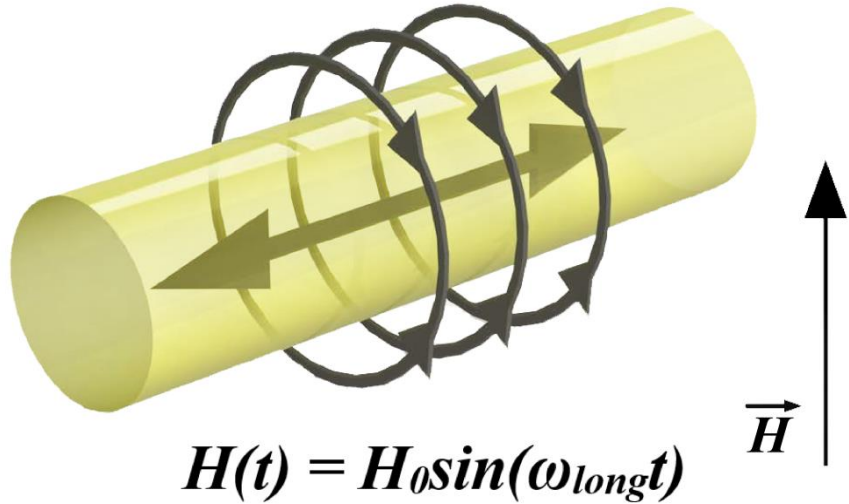

Fig. 9. Sketch of magnetic field around elongated NP (cylinder) placed into permanent magnetic field of strength H. Straight arrow inside the NP shows the oscillations of electrons in the direction of longitudinal plasmon resonance.

From the classical point of view, the magnetic field associated with longitudinal oscillations of electrons can be described as current of ultra-high frequency. The field lines associated with longitudinal oscillations are closed circles, their direction changes with frequency $\omega_{long}$. From the considerations of symmetry about one third of free electrons in Au NPs take part in these oscillations. The configuration of the fields shown in Fig. 9 is close to mechanical equilibrium – at each cycle of longitudinal oscillations the interaction of magnetic fields tends to rotate the whole NP either up or down with respect to the direction of H. However, elongated Au NPs are rather flexible than rigid. This means that if the orientation of the elongated NP does not coincide with that shown in Fig. 9, the interaction of magnetic fields will cause the deflection of either whole NP or its uniaxial part towards mechanically stable position, i.e. the long axis of the NP is perpendicular to the vector of H. In other words, if the external field is directed vertically, long axis of NPs should be oriented horizontally with arbitrary azimuth with respect to external field.

High-frequency oscillations of electrons in the magnetic field should be accompanied by significant losses of their energy on the walls of NPs. Finally this may lead to heating of NPs to high temperatures causing thus their "welding" to longer NPs. This conclusion is directly confirmed by Figs. 5 and 8, where the increase of aspect ratio of elongated NPs is clearly seen.

**Conclusion**

Thus, it is experimentally demonstrated that permanent magnetic field may influence the morphology of elongated diamagnetic Au NPs generated by laser ablation of Au target in aqueous solutions. Namely, the aspect ratio of elongated Au NPs subjected to permanent magnetic field with strength up to 7 Tesla for sufficiently long time of order of tens minutes increases from initial $6 - 8$ to $17 - 18$. The effect of elongation increases with residence time of NPs in the external magnetic field. The effect is interpreted on the basis of interaction of electrons that participate in longitudinal plasmon resonance with external field.

**Acknowledgements**

This work was performed within State Contract No. AAAA-A18-118021390190-1 and within the framework of National Research Nuclear University 'MEPhI' (Moscow Engineering Physics Institute) Academic Excellence Project (Contract No. 02.a03.21.0005) and supported in part by the Russian Foundation for Basic Research (Grant Nos 16-02-01054_a, 18-32-01044_mol_a, and 18-52-70012_e_Aziya_a), Programme No. 7 of the Presidium of the Russian Academy of Sciences. E.V. Barmina is thanked for useful discussions of the manuscript.